\documentclass[a4paper,11pt]{article}
\usepackage{pos}
\usepackage{amsmath}
\usepackage{tikz}
\usepackage{dsfont}
\usepackage[utf8]{inputenc}
\usepackage[normalem]{ulem}
\usepackage{braket}

\newcommand{\SU}{\text{SU}}

\newcommand{\nmax}{n_{\max}}
\newcommand{\Wg}{\widetilde{\text{W\!g}}}

\title{Grassmann tensor approach for two-dimensional QCD in the strong-coupling expansion}
\ShortTitle{Grassmann tensor approach for 2d QCD in the strong-coupling expansion}

\author*[a]{Thomas Samberger}
\author[a]{Jacques Bloch}
\author{Robert Lohmayer}

\affiliation[a]{Institute for Theoretical Physics, University of Regensburg, Regensburg, Germany}

\emailAdd{thomas.samberger@ur.de}
\emailAdd{jacques.bloch@ur.de}
\emailAdd{robert.lohmayer@ur.de}

\abstract{
	We present a tensor-network approach for the strong-coupling expansion of two-dimensional QCD with staggered quarks at non-zero chemical potential. After expanding the Boltzmann factor in the gauge and fermion actions, all gauge fields can be integrated out exactly and the partition function can be evaluated using the Grassmann higher-order tensor renormalization group approach. The method is modified to compute the $\mu$ dependence of the quark number density and the chiral condensate up to order $\beta^3$ with complete absence of higher-order terms infiltrating the result. Although the expansion itself is only a good approximation to the full theory at small $\beta<0.1$, the range can be extended, by using judiciously chosen fits. Moreover, these fits also yield a valuable expansion in $\beta$ for the critical chemical potential.
}

\FullConference{The 41st International Symposium on Lattice Field Theory (LATTICE2024)\\
 28 July - 3 August 2024\\
Liverpool, UK\\}

\begin{document}
	
	\maketitle
	                   
	\section{Introduction}
	For non-zero chemical potential, the sign problem prohibits the simulation of lattice QCD (LQCD) using traditional Monte Carlo methods. Therefore, we will apply a tensor-network approach based on singular-value decompositions to compute the partition function and thermodynamic observables of two-dimensional lattice QCD in the strong-coupling expansion for general orders in the coupling parameter $\beta$. While we restrict ourselves to the case of single-flavored staggered quarks, the generalization to an arbitrary number of flavors is straightforward. Additionally, we will show results for the quark number density and the chiral condensate up to order three. In previous work we studied the infinite-coupling case up to four dimensions \cite{Bloch:2022vqz,Bloch:2022yiq,Milde2023}. In the following, we will outline the tensor-network formalism for the two-dimensional strong-coupling expansion beyond leading order. For full details we refer to a forthcoming publication \cite{Samberger2025}.
		
	\section{Tensor-network approach in two dimensions}
	We start with a brief summary of the higher-order tensor renormalization group (HOTRG) approach in two dimensions, first introduced in \cite{Xie_2012}. This approach can be used whenever the partition function has the form 
	\begin{equation}
		Z=\sum_{\{j_{x,\nu}\}}\prod\limits_{x=1}^V \mathcal{T}_{j_{x,-1},j_{x,1},j_{x,-2},j_{x,2}}\label{eq:Z}
	\end{equation}
	on a lattice with $V$ sites. There is one single tensor $\mathcal{T}$, which is placed on every site $x$ of the lattice. For every lattice link\footnote{$\nu$ distinguishes the directions on the lattice and therefore $\nu\in\{1,2\}$ for the two-dimensional case. Throughout the paper we will also use the convention $(x,-\nu) \equiv (x-\hat\nu,\nu)$.} $(x,\nu)$ there is an index $j_{x,\nu}\in\{1,\dots,D_0\}$, with initial bond dimension $D_0$, which depends on the theory under consideration. Adjacent tensors are connected via contraction, i.e., the summation over a common link index $j_{x,\nu}$.

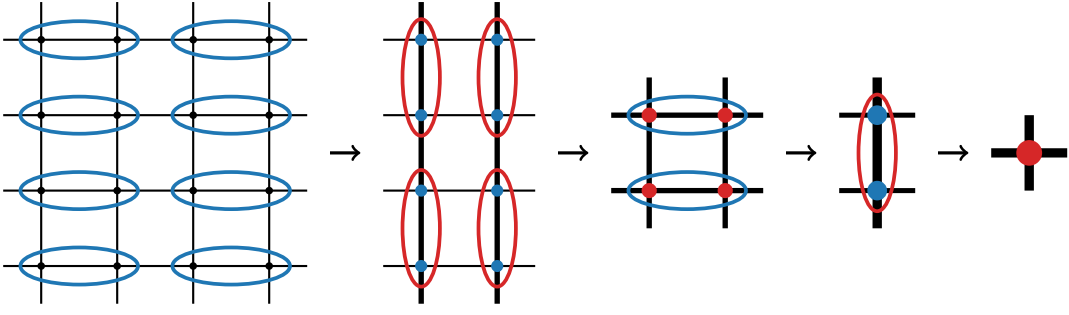
\begin{figure}[t]
		\definecolor{myblue}{rgb}{0.1216, 0.4667, 0.7059} 
		\definecolor{myred}{rgb}{ 0.8392, 0.1529, 0.1569} 
		\centering
		\begin{tikzpicture}
			\foreach \a / \b in {0/0, 1/0, 2/0, 3/0}
			{
				\draw[-, line width=0.8pt] (\a,\b-0.5) -- (\a,\b+3.5);
			}
			\foreach \a / \b in {0/0, 0/1, 0/2, 0/3}
			{
				\draw[-, line width=0.8pt] (\a-0.5,\b) -- (\a+3.5,\b);
			}

			\foreach \a in {0, 1, 2, 3}
			{
				\foreach \b in {0, 1, 2, 3}
				{
					\fill (\a,\b) circle [radius=0.05];	
				}
			}
			\foreach \a in {0, 2}
			{
				\foreach \b in {0, 1, 2, 3}
				{
					\draw[myblue,line width=1.4pt] (\a+0.5,\b) ellipse (22pt and 7pt);
				}
			}
			\draw[->, line width=1.2pt] (3.8,1.5) -- (4.2,1.5);
				
			\foreach \a / \b in {5/0, 6/0}
			{
				\draw[-, line width=2pt] (\a,\b-0.5) -- (\a,\b+3.5);
			}
			\foreach \a / \b in {5/0, 5/1, 5/2, 5/3}
			{
				\draw[-, line width=0.8pt] (\a-0.5,\b) -- (\a+1.5,\b);
			}
			
			\foreach \a in {5, 6}
			{
				\foreach \b in {0, 1, 2, 3}
				{
					\fill[myblue] (\a,\b) circle [radius=0.08];	
				}
			}
			\foreach \a in {5, 6}
			{
				\foreach \b in {0, 2}
				{
					\draw[myred,line width=1.4pt] (\a,\b+0.5) ellipse (7pt and 22pt);
				}
			}
			\draw[->, line width=1.2pt] (6.8,1.5) -- (7.2,1.5);
			
			\foreach \a / \b in {8/0, 9/0}
			{
				\draw[-, line width=2pt] (\a,\b+0.5) -- (\a,\b+2.5);
			}
			\foreach \a / \b in {8/0, 8/1}
			{
				\draw[-, line width=2pt] (\a-0.5,\b+1) -- (\a+1.5,\b+1);
			}
			
			\foreach \a in {8, 9}
			{
				\foreach \b in {1, 2}
				{
					\fill[myred] (\a,\b) circle [radius=0.1];	
				}
			}
			\foreach \b in {1, 2}
			{
				\draw[myblue,line width=1.4pt] (8.5,\b) ellipse (22pt and 7pt);
			}
			
			\draw[->, line width=1.2pt] (9.8,1.5) -- (10.2,1.5);
			
			\draw[-, line width=3.5pt] (11,0.5) -- (11,2.5);
			\draw[-, line width=1.9pt] (10.5,1) -- (11.5,1);
			\draw[-, line width=1.9pt] (10.5,2) -- (11.5,2);
			\fill[myblue] (11,1) circle [radius=0.13];
			\fill[myblue] (11,2) circle [radius=0.13];
			\draw[myred,line width=1.4pt] (11,1.5) ellipse (7pt and 22pt);	
			
			\draw[->, line width=1.2pt] (11.8,1.5) -- (12.2,1.5);
			
			\draw[-, line width=3.5pt] (13,1) -- (13,2);
			\draw[-, line width=3.5pt] (12.5,1.5) -- (13.5,1.5);
			\fill[myred] (13,1.5) circle [radius=0.17];
		\end{tikzpicture}
	
		\caption{Schematic contraction procedure for a two-dimensional tensor network. Each contraction step results in a new tensor network defined on a coarse lattice with half of the number of sites than before, but increased bond dimension of the tensor.}
		\label{fig:TN_contraction}
	\end{figure}

	The partition function can be computed efficiently by pairwise contractions of tensors, resulting in new tensors defined on a coarse lattice with half the number of sites.\footnote{Note that this contraction has to be computed only once for each renormalization group step, as the same tensor is present on every site of the lattice.} The result then has again the structure of a tensor network and the pairwise contractions can be repeated in principle. The schematic contraction procedure is visualized in Fig.~\ref{fig:TN_contraction}. However, the coarse tensor has increased bond dimension after each step, which makes an exact numerical computation unfeasible already after a few contraction steps. Therefore, HOTRG also provides a truncation scheme, which reduces the range of a ``fat index''  to a chosen bond dimension $D$ in each contraction step, based on singular-value decompositions of unfoldings of the coarse tensor \cite{DeLathauwer2000}.

	\section{LQCD partition function as tensor network}
	Here, we want to apply the HOTRG approach to the partition function of two-dimensional lattice QCD, given by
	\begin{align}\label{eq:Z_init}
		Z_{\text{QCD}}=\int \left[\prod\limits_x d\psi_x d\bar\psi_x\right]\left[\prod\limits_{x,\nu} dU_{x,\nu}\right]\left[\prod\limits_{x,\nu}{\rm e}^{S_{x,\nu}^{\mathrm{f}}}{\rm e}^{S_{x,\nu}^{\mathrm{b}}}\right]\left[\prod_{x,\mu,\nu}^{\mu\neq\nu}{\rm e}^{S^\text{G}_{x,\mu,\nu}}\right]{\rm e}^{S_\text{M}}
	\end{align}
	with gauge fields $U_{x,\nu}\in \SU(N_c)$ and fermion fields $\psi_x$ and $\bar \psi_x$, which have $N_c$ Grassmann-valued components each. The corresponding differentials are the Haar integration measure $dU_{x,\nu}$ and the Grassmann integration measure
	$d\psi_x d\bar\psi_x\equiv d\psi_{x,N_c} \cdots d\psi_{x,1} d\bar\psi_{x,1} \cdots d\bar\psi_{x,N_c}$.
	
	We use staggered fermions with forward and backward hopping terms
	\begin{equation}\label{eq:ferm_act}
		S_{x,\nu}^{\mathrm{f}}=\eta_{x,\nu}\bar{\psi}_x{\rm e}^{\mu\delta_{\nu,1}}U_{x,\nu}\psi_{x+\hat\nu}
		\qquad\text{and}\qquad
		S_{x,\nu}^{\mathrm{b}}=-\eta_{x,\nu}\bar{\psi}_{x+\hat\nu}{\rm e}^{-\mu\delta_{\nu,1}}U_{x,\nu}^\dagger\psi_x
	\end{equation}
	with chemical potential $\mu$ and the usual staggered phases $\eta_{x,\nu}$. The Wilson plaquette action and the mass term are given by
	\begin{equation}\label{eq:gauge_act}
		S^{\text{G}}_{x,\mu,\nu}=\frac \beta{2N_c}{\rm tr}\left[ U_{x,\mu}U_{x+\hat\mu,\nu}U^\dagger_{x+\hat\nu,\mu}U^\dagger_{x,\nu}\right]\qquad\text{and}\qquad S_\text{M}=2m\sum\limits_x\bar\psi_x\psi_x
	\end{equation}
	with coupling parameter $\beta$ and quark mass $m$.
	
	For the tensor-network formulation, it is essential to expand all the exponentials present in \eqref{eq:Z_init}. All the sums can be brought to the very front of the expression, where they form the contractions over the tensor-network indices.\footnote{Note that the summation indices corresponding to the Wilson plaquette action are defined on plaquettes in contrast to tensor-network indices, which are defined on links. However, they can be written in terms of link indices by rewriting each plaquette index as four link indices and introducing Kronecker deltas, which demand that all four new link indices coincide.}
	
	The remaining task is to rewrite the partition function in terms of local tensors and to compute the gauge and Grassmann integrals. To this end, we write all color contractions present in \eqref{eq:ferm_act} and \eqref{eq:gauge_act} explicitly, which allows us to move all gauge fields freely inside the partition function as long as a unique color-index symbol is chosen for every color contraction. Therefore, we can collect all gauge fields corresponding to the same link, resulting in the integral
	\begin{align}
		\int\limits_{\SU(N_c)}\hspace{-0.8em}dU\,U_{i_1j_1}\cdots U_{i_aj_a}U^\dagger_{l_1m_1}\cdots U^\dagger_{l_pm_p}.
	\end{align}
	An efficient solution of this integral was presented by Borisenko et al.\ \cite{Borisenko:2018csw} and by Gagliardi and Unger \cite{Gagliardi:2019cpa}. The result is only non-zero when $\frac{|a-p|}{N_c}\equiv q\in\mathds{N}_0$, and is given by (for $a\geq p$)\footnote{In the case $a<p$, one can use the property $\int dUf(U)=\int dUf(U^{-1})$ to obtain again an integral with $a\geq p$.}
	\begin{align}\label{eq:group_int}
			\int\limits_{\SU(N_c)}\hspace{-0.8em}dU\,U_{i_1j_1}\cdots U_{i_aj_a}U^\dagger_{l_1m_1}\cdots U^\dagger_{l_pm_p}\propto\sum\limits_{(\alpha,\gamma)\in A}\sum\limits_{\pi,\sigma\in S_p}\varepsilon_{i_{\alpha}}^{\otimes q}\delta^{m_\pi}_{i_{\gamma}} \Wg_{N_c}^{q,p}\!(\pi\cdot\sigma^{-1})\,\varepsilon^{\otimes q}_{j_{\alpha}}\delta_{j_{\gamma}}^{l_\sigma}.
	\end{align}
	Here, $\Wg$ are the so-called generalized Weingarten functions. The sum over $(\alpha,\gamma)\in A$ runs over all possible ways to partition a set of size $qN_c+p$ in $q$ subsets of size $N_c$ and one subset of size $p$, referred to by $\alpha$ and $\gamma$, respectively. As the summation variables in \eqref{eq:group_int} are defined on every link, they can be treated as additional tensor-network indices.\footnote{Due to the Grassmann antisymmetry, parts of this sum can be calculated locally yielding a smaller initial bond dimension, which is advantageous for numerical calculations.}
	
	While all gauge fields can be integrated out exactly, in a form suitable for the tensor-network formulation, the Grassmann variables cannot be integrated out without producing non-local signs. However, this issue can be resolved by introducing a Grassmann network \cite{Shimizu:2014uva}, where all original Grassmann variables, representing the quarks, are integrated out and one is left with a single auxiliary Grassmann variable on every link of the lattice, possibly yielding Grassmann-valued tensor entries and requiring a modified HOTRG algorithm called Grassmann HOTRG (GHOTRG).
	
	Using these approaches for the integrals, all color indices can be contracted locally, and therefore, they form no degrees of freedom of the tensor network \cite{Samberger2025}. This is a crucial result, as otherwise the initial bond dimension would be much larger.  
	
	\section{\boldmath Separation of different orders in $\beta$}
	\label{Sec:Separation}
	\begin{figure}[t]
		\centering
		\begin{tikzpicture}[scale=1.3]
			\foreach \a / \b in {0/0, 1/0, 2/0, 3/0}
			{
				\draw[-, line width=0.8pt] (\a,\b-0.5) -- (\a,\b+3.5);
			}
			\foreach \a / \b in {0/0, 0/1, 0/2, 0/3}
			{
				\draw[-, line width=0.8pt] (\a-0.5,\b) -- (\a+3.5,\b);
			}
			
			\foreach \a in {0, 1, 2, 3}
			{
				\foreach \b in {0, 1, 2, 3}
				{
					\fill (\a,\b) circle [radius=0.05];	
				}
			}
			\foreach \b in {0, 1}
			{
				\foreach \a in {1.7, 2.7}
				{
					\node at (\b+0.25, \a+0){$1+$};
			
					\draw[orange, -, line width=0.8pt] (\b+0.55,\a-0.162) -- (\b+0.55,\a+0.162);
					\draw[orange, -, line width=0.8pt] (\b+0.85,\a-0.162) -- (\b+0.85,\a+0.162);
					\draw[orange, -, line width=0.8pt] (\b+0.55,\a-0.15) -- (\b+0.85,\a-0.15);
					\draw[orange, -, line width=0.8pt] (\b+0.55,\a+0.15) -- (\b+0.85,\a+0.15);
					\node at (\b+0.23, \a-0.5){$+$};
					\draw[orange, -, line width=0.8pt] (\b+0.4,\a-0.662) -- (\b+0.4,\a-0.338);
					\draw[orange, -, line width=0.8pt] (\b+0.7,\a-0.662) -- (\b+0.7,\a-0.338);
					\draw[orange, -, line width=0.8pt] (\b+0.4,\a-0.65) -- (\b+0.7,\a-0.65);
					\draw[orange, -, line width=0.8pt] (\b+0.4,\a-0.35) -- (\b+0.7,\a-0.35);
		
					\draw[orange, -, line width=0.8pt] (\b+0.45,\a-0.612) -- (\b+0.45,\a-0.388);
					\draw[orange, -, line width=0.8pt] (\b+0.65,\a-0.612) -- (\b+0.65,\a-0.388);
					\draw[orange, -, line width=0.8pt] (\b+0.45,\a-0.6) -- (\b+0.65,\a-0.6);
					\draw[orange, -, line width=0.8pt] (\b+0.45,\a-0.4) -- (\b+0.65,\a-0.4);
				}	
			}
			\fill (0.5,0.4) circle [radius=0.03];
			\fill (0.5,0.5) circle [radius=0.03];
			\fill (0.5,0.6) circle [radius=0.03];	
			
			\fill (2.4,2.5) circle [radius=0.03];
			\fill (2.5,2.5) circle [radius=0.03];
			\fill (2.6,2.5) circle [radius=0.03];
		\end{tikzpicture}
		\caption{Visualization of all contributions using the example $\nmax=2$. Even though no plaquette has a contribution higher than $\nmax=2$, the partition function contains terms of much higher order, since the expansion is done on every plaquette separately. These terms are avoided in our modified GHOTRG procedure.}
		\label{fig:separation}
	\end{figure}
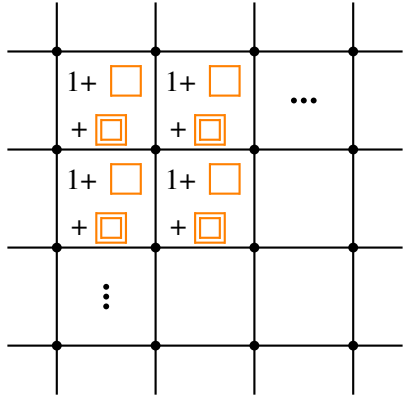

	In order to apply the GHOTRG method, we truncate the Taylor expansion of the gauge action at order $\nmax$ for every plaquette separately (visualized in Fig.~\ref{fig:separation}). A direct application of this method generates many incomplete terms of order higher than $\nmax$. Determining the first $\nmax$ coefficients by fitting the results as functions of $\beta$ turned out to produce large errors. 

	Therefore, we developed a modified GHOTRG procedure \cite{Samberger2025}, where terms contributing to orders higher than $\nmax$ are avoided in each coarsening step. Specifically, we group link indices according to the occupation of adjacent plaquettes and perform separate SVD truncations on these groupings in order to avoid a mixing of different orders. Furthermore, plaquettes get contracted into links, which demands additional index groupings. Later on, those links get contracted into sites as well, which can be distinguished by introducing different tensors on every site, depending on the site occupation. This method allows for a direct calculation of all coefficients of the strong-coupling expansion of $Z_\text{QCD}$ up to order $\nmax$ simultaneously.

	\section{Translational invariance}
	Translational invariance is a symmetry of the system that can be used to reduce the numerical complexity considerably: As many contributions differ only in the choice of the origin, we have to consider only one of those contributions and introduce a combinatorial factor to keep the result unchanged. For example, in a calculation with $\nmax=1$, we get the correct result by allowing a plaquette occupation only in the origin of the lattice and applying a factor $V$ to those contributions. Similar conditions can be found for $\nmax>1$ as well. For tensor networks, those situations can be considered by introducing four special tensors around the plaquette at the origin and tensors that are identical to each other on all other sites. After a contraction in both directions, the procedure yields an ``impurity system'', consisting of one impurity tensor and identical tensors on all remaining sites.

	\section{Results}	
	\subsection{Quark number density}
	\label{sec:qnd}
\begin{figure}[t]
		\centering
		\includegraphics[width=10cm]{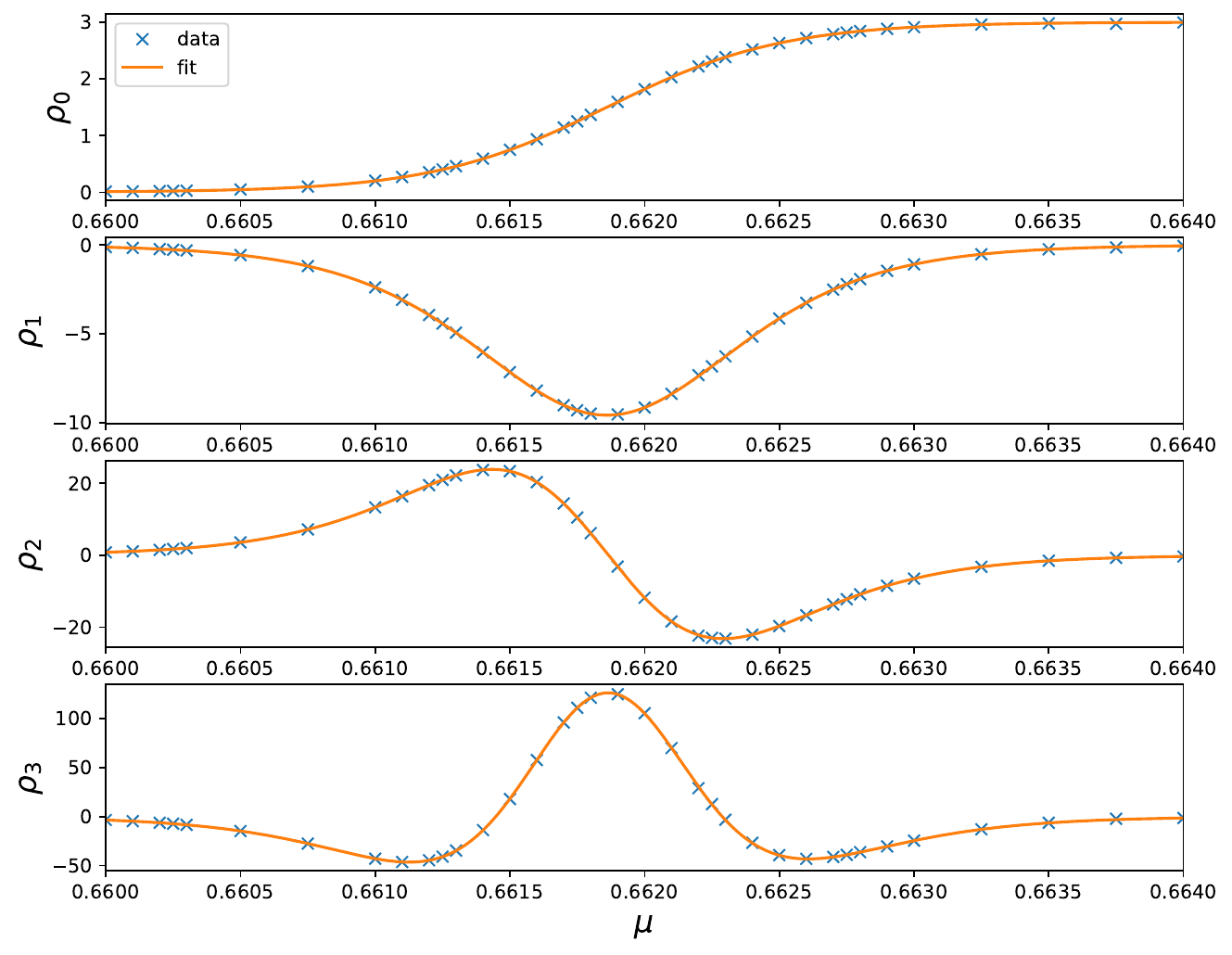}
		\caption{Results for the coefficients $\rho_n(\mu)$ of the quark number density in the strong-coupling expansion up to order three. We observe excellent agreement between tensor data (blue crosses) and the simultaneous fit (solid lines).} 
		\label{fig:baryon_density}
	\end{figure}

	We first show results for the expansion of the quark number density
	\begin{equation}\label{eq:baryon_dens_expansion}
		\rho(\beta,\mu)\equiv\frac 1V\frac{\partial\log Z(\beta,\mu)}{\partial \mu}=\sum\limits_{n=0}^{\nmax}\rho_n(\mu)\beta^n+\mathcal{O}(\beta^{\nmax+1}),
	\end{equation}
	where the method described in Sec.~\ref{Sec:Separation} allows for a direct calculation of all coefficients $\rho_n(\mu)$, by expanding the logarithm in $\beta$. The results obtained with $\nmax=3$ are shown in Fig.~\ref{fig:baryon_density} for a $32\times32$ lattice, quark mass $m=0.5$ and bond dimension $D=135$. The coefficients are consistent with those obtained with $\nmax<3$, where convergence is reached for smaller values of $D$.

	In order to model the behavior of these coefficients, we use an ansatz motivated by the Fermi-Dirac statistics,\footnote{An even better fit ansatz would be obtained using $\rho(\beta,\mu) - \rho(\beta,-\mu)$, which is antisymmetric in $\mu$. However, for $\mu \geq 0$, the additional term is negligible for the chosen quark mass and lattice size.}
	\begin{equation}\label{eq:fit-ansatz}
		\rho(\beta,\mu)=\frac32\bigg(1+\tanh[a_\rho(\beta)(\mu-\mu^{\text{c}}_\rho(\beta))]\bigg),
	\end{equation}
	which depends on a critical chemical potential $\mu^{\text{c}}_\rho(\beta)$  and a transition sharpness $a_\rho(\beta)$.
 	Using 
	\begin{align}\label{expansions}
		\mu^{\text{c}}_\rho(\beta)=\sum\limits_{n=0}^{\nmax}\mu^{\text{c}}_{\rho,n}\beta^n+\mathcal{O}(\beta^{\nmax+1})
		\qquad\text{and}\qquad
		a_\rho(\beta)=\sum\limits_{n=0}^{\nmax}a_{\rho,n}\,\beta^n+\mathcal{O}(\beta^{\nmax+1}) ,
	\end{align}
we expand the ansatz \eqref{eq:fit-ansatz} in $\beta$ to obtain fit functions for the coefficients $\rho_n(\mu)$ of the strong-coupling expansion \eqref{eq:baryon_dens_expansion} with fit parameters $\{\mu^{\text{c}}_{\rho,n},a_{\rho,n}\}_{0\leq n\leq\nmax}$. All coefficients $\rho_n(\mu)$ are fitted simultaneously, assuming an error of $1\%$ of the maximal value of $|\rho_n(\mu)|$ for each data point of $\rho_n(\mu)$. As shown in Fig.~\ref{fig:baryon_density},  we find very good agreement between fit functions and tensor data.

\begin{figure}[t]
		\centering
		\includegraphics[width=10cm]{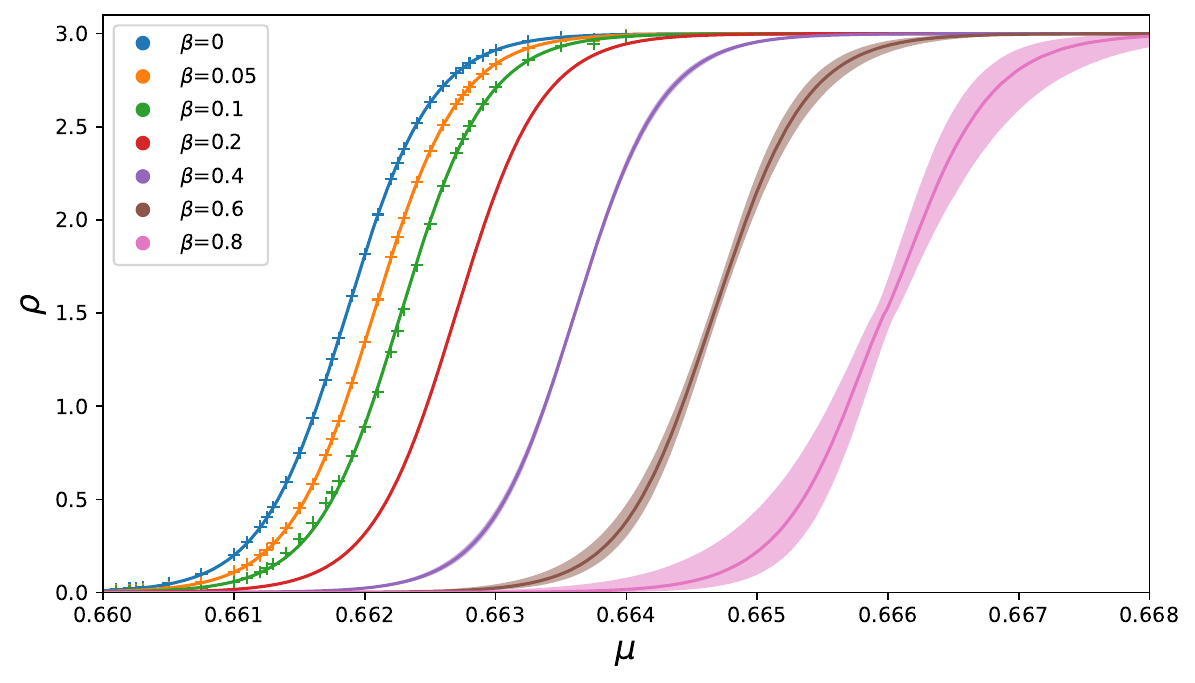}\\
		\includegraphics[width=10cm]{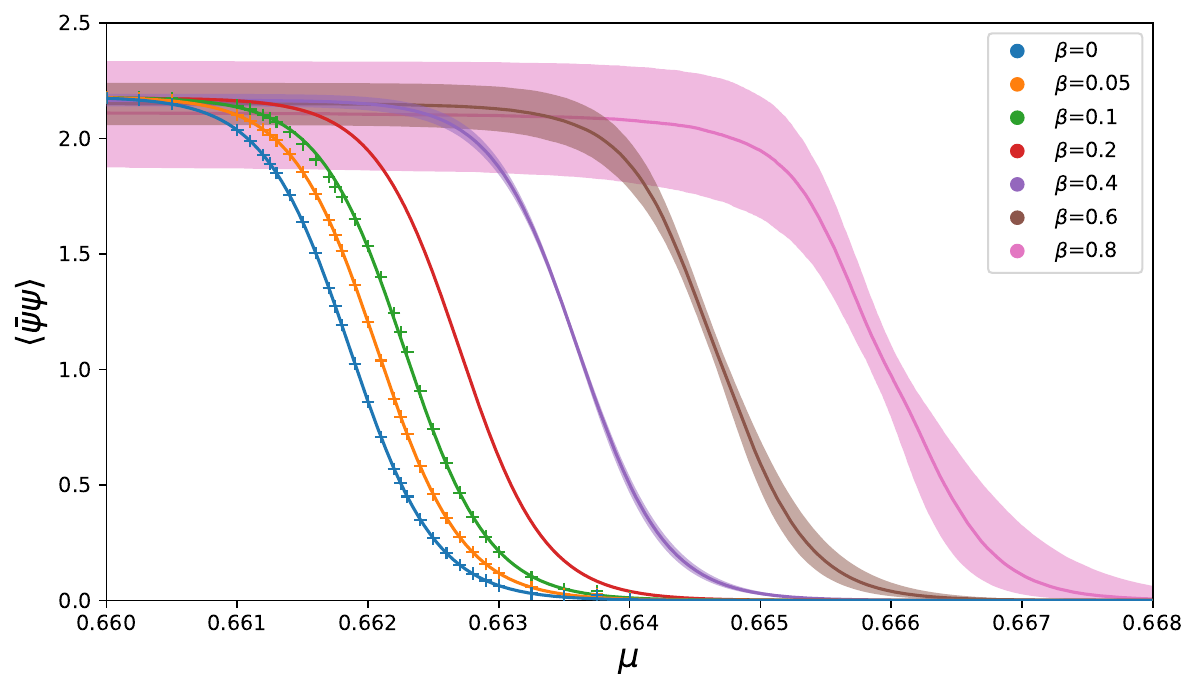}
		\caption{Quark number density (top) and chiral condensate (bottom), up to order three, as a function of $\mu$ for $m=0.5$, $V=32\times 32$, bond dimension $D=135$, and several $\beta$ values. The solid lines show the medians of the distributions of $\rho(\beta,\mu)$ and $\braket{\bar\psi\psi}(\beta,\mu)$. The error bands indicate the range between the lower and upper quartiles of these distributions. In addition, the data points show the tensor results computed directly from \eqref{eq:baryon_dens_expansion} and \eqref{eq:cc_expansion} with $\nmax=3$ for small values of $\beta\leq 0.1$. In both plots, we observe an increasing critical chemical potential with increasing coupling parameter $\beta$. }
		\label{fig:full_baryon_density}
	\end{figure}

	From the simultaneous fit we obtain the averages for $\mu^{\text{c}}_{\rho,n}$ and $a_{\rho,n}$ together with the $(2\nmax+2) \times (2\nmax+2)$ covariance matrix. For any value of $\beta$, this allows us to compute the averages of $\mu^{\text{c}}_\rho(\beta)$ and $a_\rho(\beta)$, together with the corresponding $2\times 2$ covariance matrix. Assuming a joint Gaussian distribution for $(\mu^{\text{c}}_\rho(\beta),a_\rho(\beta))$ and using \eqref{eq:fit-ansatz}, we then obtain the distribution of $\rho(\beta, \mu)$ by sampling, for fixed $\beta$ and $\mu$. In Fig.~\ref{fig:full_baryon_density} (top) we plot the results for $\rho(\beta,\mu)$ as a function of $\mu$, close to the phase transition, for several values of $\beta$.

	For small values of $\beta\leq0.1$, the expansion \eqref{eq:baryon_dens_expansion} coincides with the distribution $\rho(\beta,\mu)$. However, for larger values of $\beta$, we observe that the expansion \eqref{eq:baryon_dens_expansion} exhibits increasing unphysical behavior (not shown in the figure). Using the ansatz \eqref{eq:fit-ansatz} with the expansions \eqref{expansions}, we expect reliable results even beyond $\beta=0.1$.

	\subsection{Chiral condensate}
	Next, we show results for the chiral condensate, defined as
	\begin{equation}\label{eq:cc_expansion}
		\braket{\bar\psi\psi}(\beta,\mu)\equiv\frac 1V\frac{\partial\log Z(\beta,\mu)}{\partial m}=\sum\limits_{n=0}^{\nmax}\braket{\bar\psi\psi}_n(\mu)\beta^n+\mathcal{O}(\beta^{\nmax+1}).
	\end{equation}
	The coefficients $\langle\bar\psi\psi\rangle_n(\mu)$ can be computed analogously to those of the quark number density. For the chiral transition, we use the fit ansatz\footnote{An improved fit ansatz would be obtained using $\braket{\bar\psi\psi}(\beta,\mu)+\braket{\bar\psi\psi}(\beta,-\mu) - 2b_\text{cc}(\beta)$, which is symmetric in $\mu$.}
	\begin{equation}\label{ccfit}
	\braket{\bar\psi\psi}(\beta,\mu)=b_\text{cc}(\beta)\bigg(1-\tanh[a_\text{cc}(\beta)(\mu-\mu^{\text{c}}_\text{cc}(\beta))]\bigg).
	\end{equation}
	The distribution of $\braket{\bar\psi\psi}(\beta, \mu)$ is constructed in complete analogy to the distribution of $\rho(\beta, \mu)$, described above in Sec.~\ref{sec:qnd}, now assuming a three-dimensional Gaussian distribution for the parameters $\mu^{\text{c}}_\text{cc}(\beta)$, $a_\text{cc}(\beta)$ and $b_\text{cc}(\beta)$. The results for the chiral condensate up to order three are shown in Fig.~\ref{fig:full_baryon_density} (bottom). Similarly to the quark number density, we expect reliable results when applying the fit ansatz even for $\beta>0.1$.

	\subsection{\boldmath $\beta$ dependence of transition parameters}
\begin{figure}[t]
		\centering
		\includegraphics[width=14cm]{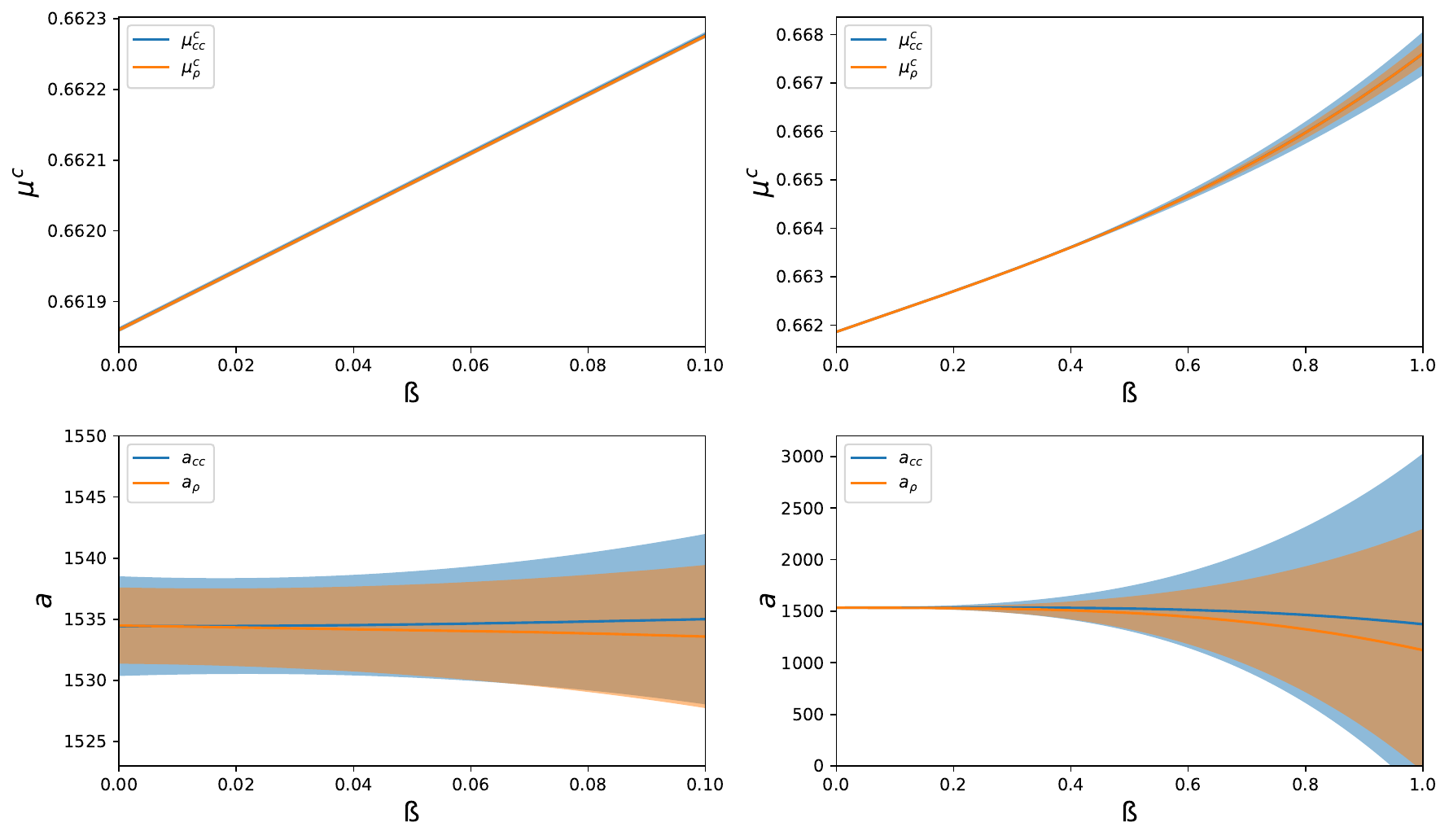}
		\caption{Critical chemical potential and transition sharpness versus coupling parameter $\beta$ for the transitions in the quark number density and the chiral condensate. The plots on the left show a close range $0\leq\beta\leq0.1$, while those on the right show a wider range up to $\beta=1$.}
		\label{fig:coeff}
	\end{figure}

	Finally, we present results for the $\beta$ dependence of the critical chemical potential, as well as the corresponding transition sharpness for both transitions in Fig.~\ref{fig:coeff}. The plots show $\sum\limits_{n=0}^{3}a_{n}\beta^n$ and $\sum\limits_{n=0}^{3}\mu^{\text{c}}_{n}\beta^n$ using the coefficients determined above. The errors on the parameters are computed as described in Sec.~\ref{sec:qnd}. We observe that the values for the critical chemical potential for both transitions coincide within the errors and that the second- and third-order terms provide only small corrections compared to the zeroth- and first-order terms. The right of Fig.~\ref{fig:coeff} shows the same curves for a wider range in $\beta$. Although we observe increasing errors on the parameters when $\beta$ is increased, the critical chemical potential remains quite accurate.
	
	\section{Conclusion}
	We have rewritten the partition function of LQCD in two dimensions as a tensor network using the strong-coupling expansion. In this approach, all color degrees of freedom are eliminated before GHOTRG is applied, which reduces the initial bond dimension. Furthermore, the standard GHOTRG procedure is modified to allow for the direct computation of all coefficients of the strong-coupling expansion simultaneously, up to a given order. We have shown results for the quark number density and chiral condensate, where both observed transitions can be very accurately modeled using the hyperbolic tangent. Future work will address the straightforward generalization to an arbitrary number of flavors, as well as a procedure for four space-time dimensions.

	\bibliographystyle{JHEP}
	\bibliography{qcd_sce}
\end{document}